\documentclass[pra,aps,twocolumn]{revtex4}
\usepackage{amssymb}
\usepackage{latexsym}
\usepackage{graphicx}
\usepackage{boldtensors}
\newcommand{\Z}{{\mathbb Z}}

\newcommand{\R}{{\mathbb R}}

\def\be{\begin{equation}}
\def\ee{\end{equation}}
\def\bea{\begin{eqnarray}}
\def\eea{\end{eqnarray}}
\def\Tr{{\rm \,Tr\,}}

\def\bfz{{\bf 0}}
\def\k{{\bf k}}
\def\e{{\bf e}}
\def\g{{\bf g}}

\def\q{{\bf q}}

\def\p{{\bf p}}

\def\x{{\bf x}}

\def\xbar{{\overline{\bf x}}}

\def\P{{\bf P}}

\def\h2m{\frac{\hbar^2}{2m}}

\begin{document}
\title{
{\flushleft{\small {\rm Published in Phys. Rev. Lett. \textbf{112}, 095301 (2014)
                                    }\\}}
%\vspace{1cm}
\vspace{1cm} \large\bf Galilean invariance in confined quantum systems: Implications for spectral gaps, superfluid flow, and periodic order\\}
\author{Andr\'as S\"ut\H o\\
Institute for Solid State Physics and Optics, Wigner Research Centre for Physics,
Hungarian Academy of Sciences, P. O. Box 49, H-1525 Budapest,  Hungary}
\thispagestyle{empty}
\begin{abstract}
\noindent
Galilean invariance leaves its imprint on the energy spectrum and eigenstates of $N$ quantum particles, bosons or fermions, confined in a bounded domain. It endows the spectrum with a recurrent structure, which in capillaries or elongated traps of length $L$ and cross-section area $s_\perp$ leads to spectral gaps $n^2h^2s_\perp\rho/(2mL)$ at wave numbers $2n\pi s_\perp\rho$, where $\rho$ is the number density and $m$ is the particle mass. In zero temperature superfluids, in toroidal geometries, it causes the quantization of the flow velocity with the quantum $h/(mL)$ or that of the circulation along the toroid with the known quantum $h/m$. Adding a "friction" potential which breaks Galilean invariance, the Hamiltonian can have a superfluid ground state at low flow velocities but not above a critical velocity which may be different from the velocity of sound. In the limit of infinite $N$ and $L$, if $N/L=s_\perp\rho$ is kept fixed, translation invariance is broken, the center of mass has a periodic distribution, while superfluidity persists at low flow velocities. This conclusion holds for the Lieb-Liniger model.

\vspace{2mm}\noindent PACS: 03.75.Kk, 05.30.-d, 67.10.Fj, 67.25.dj, 67.80.bd

\end{abstract}
\maketitle
%%%%%%%%%%%%%%%%%%%%%%%%%%%%%%%%%%%%%%

Galilei's invariance principle states that the description of the internal evolution of any system of massive particles is the same in two reference frames whose relative displacement is a translation of constant velocity. A mathematical implication is the separability of the center-of-mass (COM) motion. In nonrelativistic quantum mechanics this principle fully acts in scattering problems, for a finite number of particles in free space~\cite{Mes}. In field theoretical/operator algebraic discussions of infinite quantum systems Galilean invariance (GI) is always in the background as a broken continuous symmetry~\cite{HP,Swi,SW}. Galilean transformation is crucial in Landau's theory of superfluidity, in phenomenological~\cite{Lan,MR}  and microscopic~\cite{SW,LSY,W} treatments. The invariance principle shows up in rigorous results on one-dimensional quantum systems as the Bose gas with Dirac delta interaction~\cite{Gir,LL,LL2}.

In this Letter we investigate the manifestations of GI in confined homogeneous quantum systems. We find a partial separability of the COM motion, lending a particular structure to the energy spectrum and the set of eigenstates. There is an irreducible part without separability from which the rest of the spectrum and eigenstates can be obtained by boosting only the motion of the COM. The
difference of any eigenvalue and its counterpart in the noninteracting system is a periodic function of the wave vector. In quasi-1D geometries, as capillaries or elongated traps, this leads to vanishing spectral gaps on an arithmetic sequence of wave numbers -- thus, to ground state degeneracy -- when the length of the system tends to infinity. For an application, we study the conditions of a zero temperature superfluid flow. We postulate that a pure superfluid state exhibits a fully separated free motion of the COM in a given direction, that it is the ground state in the reference frame comoving with the COM, and that its energy in the comoving frame is the same as that of the ground state at rest. In a toroid all the three conditions are met if and only if the velocity is an integer multiple of $h/(mL)$. To obtain a critical velocity, GI must be broken by adding friction. If this is vanishing at short wavelengths, the superfluid flow remains stable at low velocities, provided that the lower edge of the excitation spectrum starts linearly in the wave number, cf.~Refs.~\cite{Lan,Fey}, but becomes unstable above a critical velocity which may not be the velocity of sound. Under the same condition on the spectrum, in most ground states of quasi-1D infinite systems superfluidity coexists with a periodic distribution of the COM. Among others, a periodicity is seen on the off diagonal of a reduced density matrix, where reduction is onto a coordinate of the COM. This is consistent also with a fragmented Bose-Einstein condensation (BEC)~\cite{rem0}. A proven example is the Lieb-Liniger model~\cite{LL,LL2}.

\vspace{2pt}\noindent{\em Galilean invariance.}--\,Consider a system of $N$ identical particles of mass $m$ in $\R^d$, with or without spin. Their potential energy $U(\x_1,\ldots,\x_N)$ is a sum of $2$,...,$\ell$-body potentials, each of which is real, permutation and shift invariant. Hard-core interactions are allowed. Above one dimension $U$ may also contain a one-body term $\sum_{j=1}^N u^{(1)}(\x_j)$ where $u^{(1)}$ is shift invariant in one direction, say, along the first coordinate axis, and fast increasing perpendicular to it.

The particles are confined in a domain $\Lambda=[-\frac{L}{2},\frac{L}{2}]\times\Lambda_\perp$. Here $\Lambda_\perp$ is (i) $[-\frac{L_2}{2},\frac{L_2}{2}]\times\cdots\times [-\frac{L_d}{2},\frac{L_d}{2}]$ with $L\geq L_2\geq\cdots\geq L_d$, (ii) $\R^{d-1}$, if $U$ contains a suitable one-body term, and (iii) any bounded $(d-1)$-dimensional domain with a smooth boundary, e.g., a disk or an annulus if $d=3$. We discuss in detail (i). To implement Galilean transformation, $U$ must be periodized. The result is $U_\Lambda(\ldots,\x_j+L_i\e_i,\ldots)=U_\Lambda(\ldots,\x_j,\ldots)$ ($i=1,\ldots,d$, $L_1= L$).

With the notation $\p_j=-i\hbar\partial/\partial\x_j$,
\be
H^\bfz=\frac{1}{2m}\sum_{j=1}^N \p_j^2+U_\Lambda
\ee
is the energy operator on the Hilbert space ${\cal H}=L^2(\Lambda^N\setminus S_{\rm exc})$, defined with periodic boundary conditions (BC). Here $S_{\rm exc}$ is the (possibly empty) set excluded from $\Lambda^N$ by a hard-core interaction. $H^\bfz$ may be restricted to the symmetric or antisymmetric subspace of $\cal H$, eigensubspaces of the total spin, or a component of it. Also with periodic BC, on the same domain $D^\bfz\subset{\cal H}$, for any vector $\g\in(\R^d)^*$ (dual or reciprocal space) we define the Galilean transforms $\p_j^\g=\p_j+\hbar\g$, $\P^\g=\sum_j \p_j^\g$ and
\be\label{GalH}
H^\g=\frac{1}{2m}\sum_{j=1}^N(\p_j^\g)^2+U_\Lambda = H^\bfz + \frac{\hbar}{m}\g\cdot\P^\g -\frac{\hbar^2}{2m}N\g^2.
\ee
$H^\g$ and $\P^\g$ are interpreted as the energy and, respectively, momentum operators of the system in the "$\g$~frame" defined to be in uniform translation of velocity $-\hbar\g/m$ with respect to $\Lambda$. The common eigenfunctions of $H^\g$ and $\P^\g$ are independent of $\g$. We have
\bea\label{g-spectra}
H^\g\psi_{\q,n}&=&E^\g_{\q,n}\psi_{\q,n}\qquad
\P^\g\psi_{\q,n}=\hbar(\q+N\g)\psi_{\q,n}\,,\nonumber\\
E^\g_{\q,n}&=&E^\bfz_{\q,n}+(\hbar^2/2m)[N\g^2+2\g\cdot\q]
\eea
where $E^\g_{\q,n}\leq E^\g_{\q,n+1}$ $(n\geq 0)$, and
\be
\q\in\Lambda^*=\{(2n_1\pi/L_1)\e_1+\cdots+(2n_d\pi/L_d)\e_d\}_{n_1,\ldots,n_d\in\Z}
\ee

For any $\g,~\gamma\in(\R^d)^*$ let $H^\g(~\gamma)$ and $\P^\g(~\gamma)$ be, as differential operators, equal to $H^\g$ and, respectively, $\P^\g$, but defined with a twist $~\gamma$ in the BC: their domain $D(~\gamma)$ consists of the functions
\be
\psi(\x_1,\ldots,\x_N)=e^{iN~\gamma\cdot\xbar}\phi(\x_1,\ldots,\x_N),
\ee
where $\xbar=N^{-1}\sum_{j=1}^N\x_j$ (the COM), and $\phi\in D^\bfz$. Thus, for $~\gamma\in\Lambda^*$, $D(~\gamma)=D^\bfz$, $H^\g(~\gamma)=H^\g$, $\P^\g(~\gamma)=\P^\g$. Twist and boost are related through
\bea\label{gHgamma}
H^\g(~\gamma)&=&e^{iN~\gamma\cdot\xbar}H^{\g+~\gamma}\,e^{-iN~\gamma\cdot\xbar}\nonumber\\ \P^\g(~\gamma)&=&e^{iN~\gamma\cdot\xbar}\P^{\g+~\gamma}\,e^{-iN~\gamma\cdot\xbar}.
\eea
The common eigenfunctions of $H^\g(~\gamma)$ and $\P^\g(~\gamma)$ are $e^{iN~\gamma\cdot\xbar}\psi_{\q,n}$ with respective eigenvalues $E^{\g+~\gamma}_{\q,n}$ and $\hbar[\q+N(\g+~\gamma)]$. The physical relevance of the $\g$~frame is in its comoving with the COM in the states $e^{-iN\g\cdot\xbar}\phi$, where $\phi$ is shift invariant parallel to $\g$. If $~\gamma-~\gamma'\in\Lambda^*$ then $D(~\gamma)=D(~\gamma')$, so $H^\g(~\gamma)=H^\g(~\gamma')$, but for any $\g,\g'$
\be\label{G1}
H^{\g'}(~\gamma)=H^\g(~\gamma)+\frac{\hbar}{2m}(\g'-\g)\cdot\left[\P^{\g'}(~\gamma)+\P^\g(~\gamma)\right]\neq H^\g(~\gamma)
\ee
unless $\g'=\g$. If $\k\in\Lambda^*$, from Eq.~(\ref{gHgamma}) and $H^\g(~\gamma+\k)=H^\g(~\gamma)$,
\be\label{G2}
H^{\g+\k}(~\gamma)=e^{-iN\k\cdot\xbar}H^\g(~\gamma)\ e^{iN\k\cdot\xbar}.
\ee
The above equations are valid also in the quasi-1D cases (ii)-(iii). Now the boundary condition is periodic or twisted in the first coordinates, $\q\in \Lambda^*=\{(2n\pi/L)\e_1\}_{n\in\Z}$, and $\g$ and $~\gamma$ are parallel to $\e_1$.

Equation (\ref{G2}) is the most concise formulation of GI in confined quantum systems. $H^\g(~\gamma)$ and $H^{\g+\k}(~\gamma)$, although different, have the same spectrum and the same set of eigenfunctions. From this, information about the spectrum and the eigenstates of $H^\g(~\gamma)$ can be extracted.

\vspace{2pt}\noindent{\em Structure of the spectrum.}--\,It suffices to consider $~\gamma=\bfz$. Take first $\g=\bfz$. From Eq.~(\ref{G2}),
\bea
H^\bfz e^{iN\k\cdot\xbar}\psi_{\q,n}&=&E^\k_{\q,n}e^{iN\k\cdot\xbar}\psi_{\q,n}\nonumber\\
\P^\bfz e^{iN\k\cdot\xbar}\psi_{\q,n}&=&\hbar(\q+N\k)e^{iN\k\cdot\xbar}\psi_{\q,n}.
\eea
Thus, with a proper choice (see below) of the phases,
\bea\label{0-spectra}
\psi_{\q+N\k,n}&=&e^{iN\k\cdot\xbar}\psi_{\q,n}\\
E^\bfz_{\q+N\k,n}&=&E^\bfz_{\q,n}+(\hbar^2/2m)[N\k^2+2\k\cdot\q]=E^\k_{\q,n}\nonumber
\eea
for all $\q,\k\in\Lambda^*$, $n\geq 0$. Then, from Eqs.~(\ref{g-spectra}) and (\ref{0-spectra}),
\be\label{g-spectra2}
E^\g_{\q+N\k,n}=E^\g_{\q,n}+(\hbar^2/2m)[N\k^2+2\k\cdot(\q+N\g)]=E^{\g+\k}_{\q,n}
\ee
for any $\g\in(\R^d)^*$, $\q,\k\in\Lambda^*$, and $n\geq 0$. One can summarize these results as follows. Define the set of irreducible $\q$ vectors by
\be
\Lambda^*_{\rm irred}=\{\q\in\Lambda^*: -\pi N/L_i<q_i\leq\pi N/L_i\ \mbox{all $i$}\}.
\ee
Then $\Lambda^*=\Lambda^*_{\rm irred}+N\Lambda^*=\{\q+N\k\}_{\q\in\Lambda^*_{\rm irred},\k\in\Lambda^*}$. If
$
{\cal F}_{\rm irred}=\{\psi_{\q,n}\}_{\q\in\Lambda^*_{\rm irred},n\geq 0}
$
then the full set ${\cal F}$ of eigenfunctions is
\be\label{calF}
{\cal F}=\bigcup_{\k\in\Lambda^*}e^{iN\k\cdot\xbar}{\cal F}_{\rm irred}.
\ee
The phase of any $\psi\in{\cal F}_{\rm irred}$ can be freely chosen; the phase of the others is fixed by Eq.~(\ref{0-spectra}). The spectrum is
\bea
\sigma(H^\g)&=&\bigcup_{\q\in\Lambda^*_{\rm irred},n\geq 0}\left[E^\bfz_{\q,n} +(\hbar^2/2m)A^\g_\q\right] \nonumber\\
A^\g_\q&=&\left\{N(\k+\g)^2+2\q\cdot(\k+\g)\right\}_{\k\in\Lambda^*}
\eea
The interaction shows up only in the irreducible part. Let $E^{\g,{\rm free}}_{\q,n}$ denote the eigenvalues of the noninteracting system, and let
$
\Delta^\g_{nn'}(\q)=E^{\g}_{\q,n}-E^{\g,{\rm free}}_{\q,n'}.
$
Then for any $\g\in(\R^d)^*$, $n,n'\geq 0$, and $\q,\k\in\Lambda^*$,
\be\label{period}
\Delta^\g_{nn'}(\q+N\k)=\Delta^\g_{nn'}(\q).
\ee
In the discussion of superfluidity we shall suppose the independent symmetry $E^{-\g}_{-\q,n}=E^\g_{\q,n}$. This holds if $U(-\x_1,\ldots,-\x_N)=U(\x_1,\ldots,\x_N)$.

\vspace{2pt}\noindent{\em Spectral gaps from recurrence.}--\,Let $s_\perp=\rho^{-1}N/L$, the $(d-1)$-dimensional "cross-section area" [quasi-1D geometries (ii)-(iii)], and $\ell_\perp=\rho^{-1}N/L^2$, the $(d-2)$-dimensional "thickness" [quasi-2D case]. For a given $\q\in\Lambda^*$, if $\tilde{q}=|q_1|$, then
\[
\min_{\bfz\neq\k\in\Lambda^*}|E^\bfz_{\q+N\k,n}-E^\bfz_{\q,n}| \leq(\hbar^2/2m)\left|4\pi^2N/L^2-4\pi \tilde{q}/L\right|%\nonumber\\
\]
\vspace{-15pt}
\be
=\frac{2\pi\hbar^2}{m}\left\{\begin{array}{ll}
|\pi s_\perp\rho-\tilde{q}|/L,&\mbox{quasi-1D} \\
|\pi\ell_\perp\rho-\tilde{q}/L|,&\mbox{quasi-2D}.
\end{array}\right.%\nonumber\\
\ee
These are upper bounds on spectral gaps valid everywhere in the spectrum and obtained with $|\k|=2\pi/L$. In particular, for $\q=\bfz$,
\be\label{q=0}
E^\bfz_{2\pi N/L,n}-E^\bfz_{0,n}=\frac{2\pi^2\hbar^2 N}{mL^2}
=\frac{2\pi^2\hbar^2}{m}\left\{\begin{array}{ll}
s_\perp\rho/L,&\mbox{quasi-1D}\\
\ell_\perp\rho,&\mbox{quasi-2D}.
\end{array}\right.
\ee

\vspace{2pt}\noindent{\em Superfluid flow.}--\,In the symmetric subspace the ground state of $H^\bfz$ is unique (hence, it is $\psi_{\bfz,0}$) and can be taken real and nonnegative~\cite{rem1}. Let $\epsilon_\k=E^\bfz_{\k,0}-E^\bfz_{\bfz,0}$. For free bosons this tends to 0 for all $\k$ if $N$ and $L_i$ go to infinity. For $N$ fixed, $\lim_{L_i\to\infty}\epsilon_\k^{\rm free}=\frac{\hbar^2\k^2}{2mN}$. If $N$ and $L_i$ are fixed, writing $|k_i|=2\pi(Nn_i+n'_i)/L_i$ with $n_i\geq 0$ and $0\leq n'_i<N$, we have
\be\label{free}
\epsilon^{\rm free}_\k=\sum_{i=1}^d\epsilon^{\rm free}_{k_i}=\sum_{i=1}^d\frac{h^2}{2mL_i^2}(Nn_i^2+2n_in'_i+n'_i).
\ee
In the case of $\k=(2\pi/L)(Nn+n')\e_1$, the corresponding eigenfunction is, apart from normalization,
$
\psi_{\k,0}^{\rm free}=\psi_{\parallel}(x_{11},\ldots,x_{N1})\prod_{j=1}^N\varphi_0(\x_{j\perp})
$
where
\be\label{freefunction}
\psi_{\parallel}=e^{i2\pi Nn\overline{x}_1/L}
\sum_{1\leq j_1<\cdots<j_{n'}\leq N}e^{i2\pi(x_{j_11}+\cdots +x_{j_{n'}1})/L},
\ee
$\overline{x}_1$ is the first component of $\xbar$, $\x_j=x_{j1}\e_1+\x_{j\perp}$, and $\varphi_0$ is the ground state of  $-(\hbar^2/2m)\partial^2/\partial\x_{\perp}^2\,[+u^{(1)}(\x_\perp)]$ in $\Lambda_\perp$. In case (i) and periodic BC $\varphi_0\equiv 1$.

We make an implicit assumption on $U$ via $\epsilon_{\k}$, which may hold for repulsive interactions: there exists a $c=c(\rho)>0$ such that
\be\label{cond1}
\epsilon_\k- \epsilon_\k^{\rm free}\equiv\Delta^\bfz_{00}(\k)-\Delta^\bfz_{00}(\bfz)\geq 2N\hbar c \sum_{i=1}^d\frac{1}{L_i}\left|\sin\frac{L_ik_i}{2N}\right|;
\ee
in quasi-1D there is a single term. By Eq.~(\ref{period}), $\epsilon_\k- \epsilon_\k^{\rm free}$ is periodic in $\k$ with period lengths $2\pi N/L_i$ and zeros in $\k\in N\Lambda^*$ where $\epsilon_{\k}=\epsilon_\k^{\rm free}=\frac{\hbar^2\k^2}{2mN}$, cf.~Eq.~(\ref{0-spectra}). If Eq.~(\ref{cond1}) holds for $0\leq k_i\leq \pi N/L_i$, it extends to all $\k$ via $\epsilon_{-\k}=\epsilon_\k$ and Eq.~(\ref{0-spectra}). It is in this way that Eq.~(\ref{cond1}) follows from Lieb's proof for the 1D Bose gas with $\delta$ interaction~\cite{LL2}, and is consistent with (but not proved in) the theory of elementary excitations in experimental situations~\cite{Lan,Fey}.

We prove that the unique minimum of the full set of eigenvalues $\{E^\g_{\q,n}\}_{\g\in(\R^d)^*,\q\in\Lambda^*,n\geq 0}$ is $E^\bfz_{\bfz,0}$. If $\g\in\Lambda^*$ then $E^\g_{\q,n}=E^\bfz_{\q+N\g,n}>E^\bfz_{\bfz,0}$ unless $\q+N\g=\bfz$, $n=0$. For $\g$ not in $\Lambda^*$ write $\g=[\g]+\{\g\}$ where $[\g]\in\Lambda^*$ is (one of) the closest neighbor(s) of $\g$ in $\Lambda^*$. Then $E^\g_{\q,n}=E^{\{\g\}}_{\q+N[\g],n}$. Setting $\k=\q+N[\g]$, with the help of Eq.~(\ref{g-spectra}),
\be\label{Egk0}
E^{\{\g\}}_{\k,0}=E^\bfz_{\bfz,0}+N\frac{\hbar^2\{\g\}^2}{2m}+\epsilon_\k +\frac{\hbar^2}{m}\{\g\}\cdot\k.
\ee
Because $|\{\g\}\cdot\e_i|\leq\pi/L_i$,
\be\label{diff}
\epsilon_\k +\frac{\hbar^2}{m}\{\g\}\cdot\k\geq \epsilon_\k-\frac{\pi\hbar^2}{m}\sum_{i=1}^d\frac{|k_i|}{L_i} \geq \epsilon_\k^{\rm free}-\frac{\pi\hbar^2}{m}\sum_{i=1}^d\frac{|k_i|}{L_i}.
\ee
From Eq.~(\ref{free}) it then follows that the right member of Eq.~(\ref{diff}) is nonnegative for all $\k$. Hence, because $\{\g\}\neq 0$, $E^{\{\g\}}_{\k,n}\geq E^{\{\g\}}_{\k,0}>E^\bfz_{\bfz,0}$. Observe that from Eq.~(\ref{cond1}) we have used only $\epsilon_\k\geq\epsilon_\k^{\rm free}$.

Given $\g$, among the elements of ${\cal H}$ the closest to a pure superfluid state are those with a fully separable free motion parallel to $\g$ of the COM, i.e., of the form
%$\g\cdot\P^\g\psi=0$, or $\psi=
$e^{-iN\g\cdot\xbar}\phi$ where $\g\cdot\P^\bfz\phi=0$. In order to be compatible with a zero temperature spontaneous flow of velocity $-\hbar\g/m$, the energy of $e^{-iN\g\cdot\xbar}\phi$ in the comoving $\g$~frame must be the same minimum, $E^\bfz_{\bfz,0}$, as that of the fluid at rest. Therefore, the candidates are the ground states of the Hamiltonians $H^\g(~\gamma)$, describing the system in the $\g$~frame. The unique solution is $~\gamma=-\g$:
\be\label{super}
H^\g(-\g)e^{-iN\g\cdot\xbar}\psi_{\bfz,0}=E^\bfz_{\bfz,0}e^{-iN\g\cdot\xbar}\psi_{\bfz,0}.
\ee
Note that the flow velocity can vary continuously only if the boundary condition varies simultaneously. For a fixed $~\gamma$ the flow velocity is quantized, $\g\in\Lambda^*-~\gamma$. This follows from Eq.~(\ref{super}) and $H^{\k-~\gamma}(~\gamma)=H^{\k-~\gamma}(~\gamma-\k)$ for $\k\in\Lambda^*$. Specifically, in a toroidal geometry the single valuedness of the wave functions dictates periodic BC, so in a spontaneous flow, $g$ and the velocity can only be an integer multiple of $2\pi/L$ and $h/(mL)$, respectively. [For $^4$He the latter equals $10^{-3}/L$[cm] cm/s.] Integration of the velocity along the first axis yields the quantization of circulation with the known quantum $h/m$. The quantization of the velocity and circulation can be deduced from a phenomenological macroscopic wave function~\cite{Ons,Fe}; here, we derived it from first principles. Quantized persistent flow was recently measured in ultracold trapped atomic gases~\cite{R,M,W1,W2}.

\vspace{2pt}
\noindent{\em Stability and breakdown of superfluidity.}--\,We work with $~\gamma=\bfz$. Because of GI, the interaction with the container, built into the eigenstates of $H^\bfz$ via the BC, cannot slow down a however fast superfluid flow. An extra interaction -- friction -- must be added to modify the velocity dependence of $H^\g$. This amounts to breaking GI (\ref{G2}), e.g., by replacing $H^\g$ with
\be\label{Hgeta}
H^\g_\eta=H^\g-\frac{\hbar}{m}\eta\left(|\P^\g|/\hbar\right)\,\g\cdot\P^\g.
\ee
Here $\eta(k)\geq 0$ is decreasing. Its decay expresses the fact that short wavelength excitations need a larger momentum transfer and are less probable. We assume that the slope of $\eta$ at $k=2\pi N/L$ is not infinite, and $\eta(2\pi N/L)=\zeta a/L$ where $\zeta\geq 0$ and $a$ is a characteristic length. [In quasi-1D one may choose $a=(s_\perp)^{1/(d-1)}$, because the critical velocity decreases with $s_\perp$~\cite{MR}.\,]  With an $\eta$ of a slower decay the persistent flow can only be metastable at $g\gtrsim 1/L$. The breakdown of GI is seen from
\be\label{invGal}
H^\bfz_\eta=H^\bfz\neq H^\bfz-\frac{\hbar}{m}\eta(|\P^\bfz|/\hbar)\g\cdot\P^\bfz \equiv [H^\g_\eta]^{-\g},
\ee
the inverse Galilean transform of $H^\g_\eta$. The eigenstates of $H^\g_\eta$ are still those of $H^\bfz$; the energy of $\psi_{\q,n}$ is
\be
E^{\g,\eta}_{\q,n}=E^\g_{\q,n}-\frac{\hbar^2}{m}\eta\left(|\q+N \g|\right)\,\g\cdot(\q+N \g).
\ee
For $\g\in\Lambda^*$ its minimum is obtained by choosing $\k$ to minimize
\be\label{exciteta}
E^{\g,\eta}_{\k-N\g,0}
-E^\bfz_{\bfz,0} =\epsilon_\k-\frac{\hbar^2g}{m} k\eta(k).
\ee
Here $k=|\k|$, $g=|\g|$, and we have taken into account that the minimum is attained with a $\k$ parallel to $\g$. For free bosons there is no superfluidity if $g>\pi/L\eta(0)$, because the minimum is at a $k>0$ [$k=2\pi n'/L$, $g=-2\pi n/L$ in Eq.~(\ref{freefunction})]. In the interacting case, because of Eq.~(\ref{cond1}), for $g$ small but nondecaying as $L\to\infty$ the minimum of the expression~(\ref{exciteta}) is at $k=0$, making the symmetry breaking term in Eq.~(\ref{Hgeta}) inefficient and the unique ground state of $H^\g_\eta$ to be $\psi_{-N\g,0}=e^{-iN\g\cdot\xbar}\psi_{\bfz,0}$ with energy $E^\bfz_{\bfz,0}$. At some
$
g_{\rm cr}\leq\min\left\{\frac{m\tilde{c}}{\eta(0)\hbar},\frac{\pi}{\zeta a}\right\},
$
where $\hbar\tilde{c}$ is the slope of $\epsilon_\k- \epsilon_\k^{\rm free}$ at $k=0$ ($\tilde{c}\geq c$), there is level crossing. For $g>g_{\rm cr}$, $\psi_{-N\g,0}$ becomes an excited state of $H^\g_\eta$ with unchanged energy, while in the ground state $\psi_{\k-N\g,0}=e^{-iN\g\cdot\xbar}\,\psi_{\k,0}$ the $\P^\g$ eigenvalue $\hbar \k$ is nonzero and parallel to $\g$, corresponding to a dragged fluid.
The transition may set in at $k=0$ or $k=2\pi N/L$ or somewhere in between, involving phonon or single-particle or other types of excitations.

\vspace{2pt}
\noindent{\em Periodic order in quasi-1D.}--\,In the quasi-1D geometries (ii)-(iii),  if $N,L\to \infty$ with $N/L=s_\perp\rho\equiv\lambda^{-1}$ kept fixed, then $N\Lambda^*=\{2n\pi \lambda^{-1}\,\e_1\}_{n\in\Z}$, and in the limit of infinite length Eq.~(\ref{cond1}) becomes $\epsilon_\k\geq 2\hbar c \lambda^{-1}|\sin(k\lambda/2)|$ with equality at zero value if $k\in {\cal K}\equiv\{2n\pi /\lambda\}_{n\in\Z}$. We stress that $\lim\epsilon_\k=0$ for $k\in {\cal K}$ and its periodicity of period $2\pi/\lambda$ are unconditional; its positivity outside ${\cal K}$ depends on Eq.~(\ref{cond1}). It follows that for each $k\in{\cal K}$, $\psi_{\k,0}=e^{ik\overline{x}_1}\psi_{\bfz,0}$ asymptotically becomes degenerate with the ground state. Because $\psi_{\bfz,0}$ depends on $Nd-1$ coordinates independent of $\overline{x}_1$, the set $\{\psi_{\k,0}\}_{k\in{\cal K}}$ forms a basis for the Fourier expansion of $\lambda$-periodic functions of $\overline{x}_1$. This implies that in infinite volume translation invariance is broken: the distribution of $\overline{x}_1$ becomes a $\lambda$-periodic (generalized) function.

There is a distinguished ground state, obtained as the zero temperature limit of an infinite-volume thermal equilibrium state. Consider the density matrix
\be\label{mu}
\mu_{\Lambda}=Z_{\Lambda}^{-1}\sum_{\k,n}e^{-\beta(E^\bfz_{\k,n}-E^\bfz_{\bfz,0})}|\psi_{\k,n}\rangle\langle\psi_{\k,n}|.
\ee
Above, the sum is $e^{-\beta(H^\bfz-E^\bfz_{\bfz,0})}$, and $Z_{\Lambda}$ is its trace. As $N,L\to\infty$ with $N/L=\lambda^{-1}$, $E^\bfz_{\k,n}-E^\bfz_{\bfz,0}$ goes to zero if $k\in{\cal K}$ at least for $n=0$ (and possibly for any fixed $n$), while the difference remains positive for other values of $k$. We apply the change of variables $(\x_1,\ldots,\x_N)\rightarrow (\overline{x}_1,X')$ where $X'=\{x_{j1}-x_{11}\}_{j=2}^N\cup\{\x_{j\perp}\}_{j=1}^N$, take the partial trace over $X'$, and perform the two limits to obtain
\be\label{proj}
\lim_{\beta\to\infty}C(\beta)\lim_{N,L\to\infty}A(N)\Tr_{X'}\left[\mu_{\Lambda}\right]=\sum_{k\in{\cal K}}a_k|k\rangle\langle k|.
\ee
Here $\langle\overline{x}_1|k\rangle=\lambda^{-1/2} e^{ik\overline{x}_1}$, normalized to a period, and the scaling functions $C(\beta)$ and $A(N)\geq N$ are chosen according to the $n$~dependence and large-$N$ behavior of $E^\bfz_{\k,n}-E^\bfz_{\bfz,0}$. If, e.g.,~$\lim_{N,L}(E^\bfz_{\k,n}-E^\bfz_{\bfz,0})=0$ only for $k\in{\cal K}$ and $n=0$, then $C(\beta)=\lambda/\lambda_\beta$ ($\lambda_\beta$ is the thermal wavelength) and $A(N)=N^{\frac{3}{2}}$ yield $a_k\equiv 1$. If $E^\bfz_{\k,n}-E^\bfz_{\bfz,0}\sim (n+1) k^2/N+n\delta_{k,0}/N$ for $k\in{\cal K}$, then $C(\beta)\equiv 1$ and $A(N)=N$ give $a_k\sim (k^2+\delta_{k,0})^{-1}$ for $k\in{\cal K}$. The integral kernel $\lambda^{-1}\sum_{k\in{\cal K}}a_k e^{ik(\overline{x}_1-\overline{y}_1)}$ of the right member of Eq.~(\ref{proj}) is a $\lambda$-periodic (generalized) function of $\overline{x}_1-\overline{y}_1$. In the first example it is $\sum_{n=-\infty}^\infty\delta(\overline{x}_1-\overline{y}_1+n\lambda)$. Equation (\ref{proj}) may also indicate a fragmented BEC into the set ${\cal K}$~\cite{rem0}.

The periodicity of the position of the COM nicely illustrates Goldstone's theorem~\cite{Gold}, that to break a specific continuous symmetry the gap to some specific excitations must vanish. Additionally, the example of the free Bose gas illustrates the antithesis of Goldstone's theorem: undiscriminating gap vanishing does not lead to broken symmetry.

In liquid $^4$He $\lambda$ is far too small because of the high density. In dilute trapped gases $\lambda=L/N$  can be a few nanometers, which may be sufficient for the measurability of the periodic distribution.

\vspace{2pt}
\noindent{\em Coexistence of periodic order and superfluid flow in quasi-1D.}--\,A convenient way to discuss this question is to use $[H^\g_\eta]^{-\g}$, cf. Eq.~(\ref{invGal}). Choose $\g\in\Lambda^*$ nearly constant, meaning that it can vary only to stay in $\Lambda^*$ as $L$ increases. By GI, $[H^\g_\eta]^{-\g}$ is isospectral with $H^\g_\eta$. Let $E_0(H^\g_\eta)$ denote the lowest eigenvalue of $H^\g_\eta$. Write down Eq.~(\ref{mu}) with $[H^\g_\eta]^{-\g}-E_0(H^\g_\eta)$ replacing $H^\bfz-E^\bfz_{\bfz,0}$. For $g<g_{\rm cr}$, $E_0(H^\g_\eta)=E^\bfz_{\bfz,0}$, and the argument leading to Eq.~(\ref{proj}) can be repeated. For $g>g_{\rm cr}$, $E_0(H^\g_\eta)<E^\bfz_{\bfz,0}$, the energy minimum is attained at some $k\in(0,2\pi/\lambda)$, the ground state degeneracy breaks down, and both superfluidity and periodic order cease to exist.

\vspace{2pt}
\noindent
Setting $s_\perp=1$, the results of this Letter apply at all densities to the Lieb-Liniger model~\cite{LL,LL2}. The key to both superfluidity and periodic order is Eq.~(\ref{cond1}). If a system of fermions has such an excitation spectrum, the conclusions concerning superfluidity or superconductivity and long-range order are the same as for bosons.

\vspace{2pt}
\noindent
\emph{Acknowledgments.}--\,I thank P\'eter Sz\'epfalusy for a decade's countless discussions on BEC and superfluidity. This work was supported by OTKA Grant No. K109577.

\end{document}